\documentclass{article}

\usepackage{graphicx}

\begin{document}

\title{Overlap Effects on Benzene Transmission}

\date{December 7, 2015}

\author{Kenneth W. Sulston\thanks{corresponding author (sulston@upei.ca)} 
\thanks{School of Mathematical Sciences, University of Prince Edward Island, Charlottetown, PE, C1A 4P3, Canada}\and Sydney G. Davison\thanks{Department of Applied Mathematics, University of Waterloo, Waterloo, ON, N2L 3G1, Canada} \thanks{Department of Physics and the Guelph-Waterloo Physics Institute, University of Waterloo Campus, Waterloo, ON, N2L 3G1, Canada}}

\maketitle

\section{Abstract}

The H\"uckel molecular-orbital method (with overlap $S$) is used to derive the
$S$-modified version of the renormalization equations, which are then employed
to introduce overlap into the para-, meta- and ortho-benzene dimers' parameters.
Invoking the Lippmann-Schwinger scattering theory enables the spectral energy
transmission function $T(E)$ to be found for each of the benzene types.
The effect of overlap on the behaviour of the various $T(E)$ curves is, indeed,
marked, even for low values of $S$, where all the curves' symmetries become
permanently broken. As $S$ increases, the graphs become more distorted and suffer
displacements to lower energies. These results are so significant that they justify
the inclusion of overlap in all $T(E)$ studies of benzene.

\section{Introduction}

Benzene is one of the most important organic molecules, because of its structural 
simplicity and its presence as a constituent of larger compounds. Consequently, the molecule's various
properties have been extensively studied, over the years. In particular, its electron
transport properties are of interest \cite{ref1}, as being fundamental to the study
of molecular wires, of various configurations and with differing components.

Although the H\"uckel molecular-orbital method \cite{ref2} has contained the overlap
integral $S$, since its earliest inception in the last century, overlap is quite often overlooked
in quantum calculations, when its magnitude is deemed so small, as to render its 
inclusion unnecessary. A case in point arises in benzene transmission studies, where,
because overlap has been considered negligible, it has so far not appeared in any
reports (see, e.g., \cite{ref3}). Here, we  redress this situation, by introducing 
overlap into the rescaling treatment of these molecules, and discover that overlap
does, in fact, play a major role in determining the behaviour of the energy 
spectrum of the benzene transmission probability function $T(E)$.

The discrete Schr\"odinger equation gives rise to the rescaling technique \cite{ref4},
in which certain atomic sites are decimated, accompanied by renormalization of the 
parameters of the surviving sites. By this recursive process, the para-, meta- and ortho-benzenes
can be mapped onto equivalent dimer molecules \cite{ref5}. Such an approach facilitates
the application of the Lippmann-Schwinger scattering theory \cite{ref6} for
calculating the transmission-energy spectral function $T(E)$ of the renormalized
benzene molecules, thereby gaining insight into their electronic structure. In this paper,
the rescaling method is extended, in a straightforward manner, to include overlap, allowing
its significant effect on the three benzene types to be explored.

\section{Rescaling Technique including Overlap}

\begin{figure}[h]
\includegraphics[width=12cm]{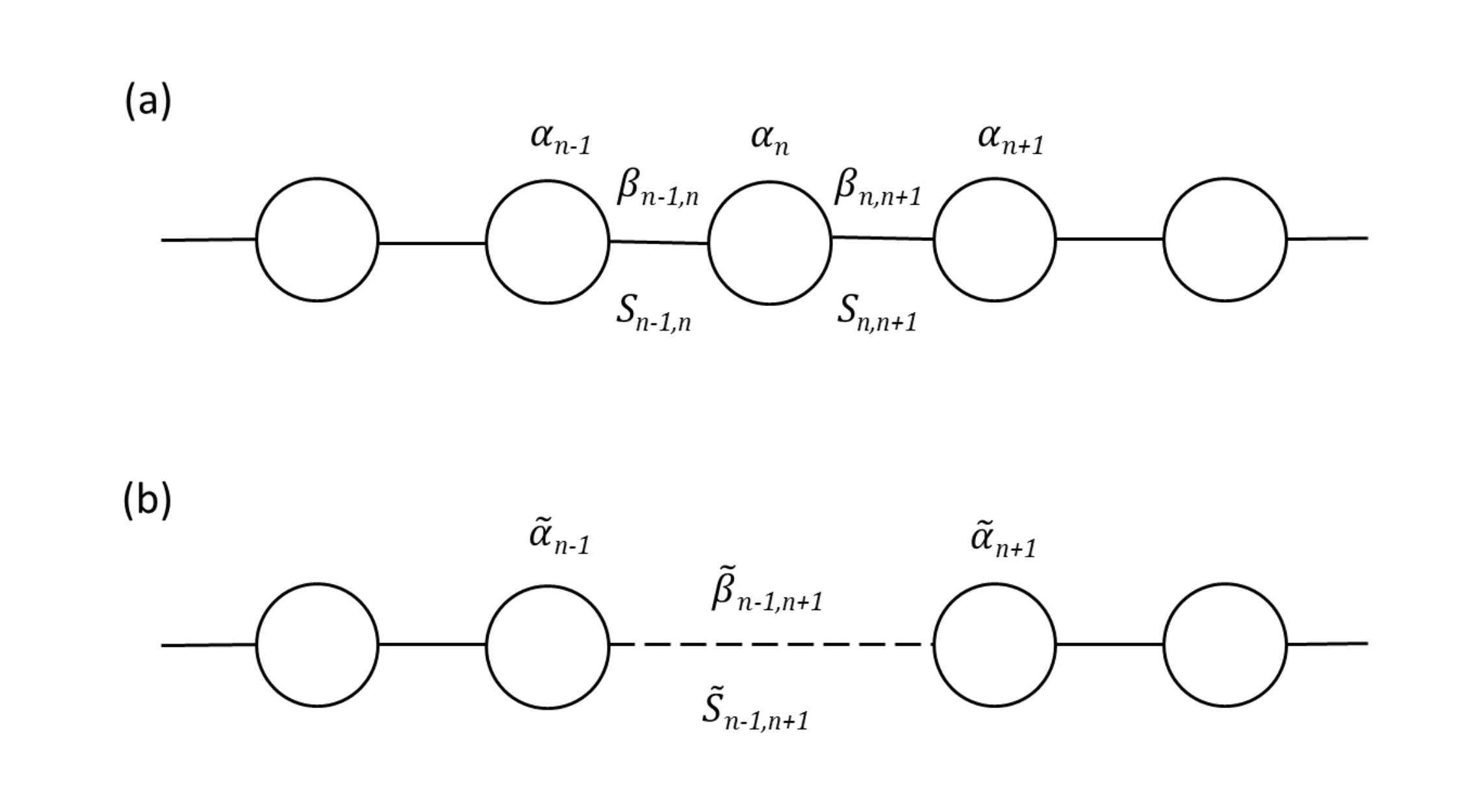}
\caption{(a) Chain in initial configuration. (b) Rescaled chain, with site $n$ decimated and and sites $n \pm 1$ renormalized.
Other sites (with parameters not shown) remain unchanged by the process.}
\label{fig1}
\end{figure}

The molecular-orbital method is applied to a linear atomic chain (see Figure \ref{fig1}(a)),
with Coulomb integral $\alpha_n$ at site $n$, bond integrals $\beta_{n-1,n}$
and $\beta_{n, n+1}$ with neighbouring sites $n-1$ and $n+1$, respectively, and corresponding
overlap integrals $S_{n-1,n}$ and $S_{n,n+1}$.
The discretized version of the Schr\"odinger equation leads to the familiar
secular equations \cite{ref2}, of which those involving site $n$ are:
\begin{equation}
(E - \alpha_{n-1})c_{n-1} = (\beta_{n-2,n-1}-ES_{n-2,n-1})c_{n-2} +  (\beta_{n-1,n}-ES_{n-1,n})c_{n},
\label{eq1}
\end{equation}
\begin{equation}
(E - \alpha_{n})c_{n} = (\beta_{n-1,n}-ES_{n-1,n})c_{n-1} +  (\beta_{n,n+1}-ES_{n,n+1})c_{n+1},
\label{eq2}
\end{equation}
\begin{equation}
(E - \alpha_{n+1})c_{n+1} = (\beta_{n,n+1}-ES_{n,n+1})c_{n} +  (\beta_{n+1,n+2}-ES_{n+1,n+2})c_{n+2}.
\label{eq3}
\end{equation}
In order to eliminate site $n$ from the set, we solve (\ref{eq2}) for $c_n$, and substitute into 
(\ref{eq1}) (and similarly into (\ref{eq3})), resulting in
\begin{eqnarray}
(E - \alpha_{n-1})  c_{n-1} = (\beta_{n-2,n-1}-ES_{n-2,n-1})c_{n-2} +  (\beta_{n-1,n}-ES_{n-1,n}) \nonumber\\  \times  \left({\beta_{n-1,n}-ES_{n-1,n} \over E - \alpha_{n}}c_{n-1}+{\beta_{n,n+1}-ES_{n,n+1} \over E - \alpha_{n}}c_{n+1} \right) ,
\label{eq4}
\end{eqnarray}
which can be rearranged as
\begin{eqnarray}
\left[ E - \alpha_{n-1}  -  {(\beta_{n-1,n}-ES_{n-1,n})^2 \over E - \alpha_{n}} \right]  c_{n-1} = (\beta_{n-2,n-1}-ES_{n-2,n-1})c_{n-2} \nonumber\\
+ { (\beta_{n-1,n}-ES_{n-1,n}) ({\beta_{n,n+1}-ES_{n,n+1}) }\over E - \alpha_{n}}c_{n+1} .
\label{eq5}
\end{eqnarray}
Equation (\ref{eq5}) can now be written in rescaled form as
\begin{equation}
( E - \tilde{\alpha}_{n-1}  )  c_{n-1} = (\beta_{n-2,n-1}-ES_{n-2,n-1})c_{n-2} 
+  (\tilde{\beta}_{n-1,n+1}-E\tilde{S}_{n-1,n+1})c_{n+1} ,
\label{eq6}
\end{equation}
where the renormalized parameters are
\begin{equation}
\tilde{\alpha}_{n-1}  = \alpha_{n-1} + {(\beta_{n-1,n}-ES_{n-1,n})^2 \over E - \alpha_{n}}  ,
\label{eq7}
\end{equation}
\begin{equation}
\tilde{\beta}_{n-1,n+1} = { (\beta_{n-1,n}-ES_{n-1,n}) ({\beta_{n,n+1}-ES_{n,n+1}) }\over E - \alpha_{n}} ,
\label{eq8}
\end{equation}
\begin{equation}
\tilde{S}_{n-1,n+1} =0.
\label{eq9}
\end{equation}
Similarly, the substitution of (\ref{eq2}) into (\ref{eq3}) leads to the rescaled equation
\begin{equation}
( E - \tilde{\alpha}_{n+1}  )  c_{n+1} = (\tilde{\beta}_{n-1,n+1}-E\tilde{S}_{n-1,n+1})c_{n-1} 
+ (\beta_{n+1,n+2}-ES_{n+1,n+2})c_{n+2}  ,
\label{eq10}
\end{equation}
where $\tilde{\alpha}_{n+1}$ is defined similarly to (\ref{eq7}). Taken together, equations
(\ref{eq6}) and (\ref{eq10}) replace (\ref{eq1})-(\ref{eq3}) in the set of secular equations,
with site $n$ having been decimated, i.e., deleted from the set, with its effect being incorporated
within the renormalized parameters (\ref{eq7})-(\ref{eq9}). The process of decimation-renormalization
can be repeated, atom by atom, until the original chain or molecule is reduced down to one, or a few,
site(s). Looking at the renormalized parameters (\ref{eq7})-(\ref{eq9}), it is important to notice that
the {\em de facto} effect of including overlap is simply to replace $\beta$  by $\beta - ES$. Consequently,
it is straightforward to extend previous work on benzene transmission so as to incorporate overlap,
as in the next section.

\section{Benzene Transmission}

In considering electron transmission through a benzene molecule, three cases arise, due to the
different atomic sites to which the molecular leads can be attached (see Figure 2 of reference \cite{ref4}.)
In the para (p)-benzene case, the leads are attached to the (1,4) sites, while in the meta (m)-benzene
and ortho (o)-benzene varieties, they are joined to the (1,5) and (1,6) sites, respectively. As shown in 
detail in \cite{ref4}, the rescaling method is applied in such a way as to successively decimate four of the
 atoms in the benzene ring, leaving a dimer consisting of the two renormalized sites attached to the
 leads (which are unaffected by the process). As indicated in the last section, the inclusion of overlap
 in the rescaling process requires only the replacement of $\beta$ by $\beta - ES$, and subsequently
 of the reduced energy $X=(E-\alpha)/\beta$ by $X_s=(E-\alpha)/(\beta - ES)$. This substitution is made
 only in the renormalized parameters, as we are only considering overlap between atoms in the benzene
 molecule, and not those in the leads. Thus, the zero-overlap
 results of \cite{ref4} are readily extended to take overlap into account. 

To this end, renormalized parameters for each type of benzene dimer, modified from the results of
\cite{ref4}, are as follows.
\newline (a) {\bf p-benzene}
\begin{equation}
\bar{\alpha}_p \equiv \tilde{\alpha}_1= \tilde{\alpha}_4 = \alpha + 2 (\beta -ES) X_s (X_s^2-1)^{-1}  ,
\label{eq11}
\end{equation}
\begin{equation}
\bar{\beta}_p \equiv \tilde{\beta}_{14} = 2 (\beta -ES) (X_s^2-1)^{-1} .
\label{eq12}
\end{equation}
\newline (b) {\bf m-benzene}
\begin{equation}
\bar{\alpha}_m \equiv \tilde{\alpha}_1= \tilde{\alpha}_5 = \alpha + (\beta -ES) X_s^{-1} [(X_s^2-1) (X_s^2-2)^{-1} +1]  ,
\label{eq13}
\end{equation}
\begin{equation}
\bar{\beta}_m \equiv \tilde{\beta}_{15} =  (\beta -ES) X_s^{-1} (X_s^2-1) (X_s^2-2)^{-1}.
\label{eq14}
\end{equation}
\newline (c) {\bf o-benzene}
\begin{equation}
\bar{\alpha}_o \equiv \tilde{\alpha}_1= \tilde{\alpha}_6 = \alpha + (\beta -ES) X_s^{-1} [1+ (X_s \bar{X}_s)^{-1}]  ,
\label{eq15}
\end{equation}
\begin{equation}
\bar{\beta}_o \equiv \tilde{\beta}_{16} =  (\beta -ES) (X_s^2-1) (X_s^2-2) [(X_s^2-1)^2-X_s^2]^{-1},
\label{eq16}
\end{equation}
with
\begin{equation}
\bar{X}_s = (X_s - X_s^{-1}) -  (X_s - X_s^{-1})^{-1} .
\label{eq17}
\end{equation}

Also adapted from \cite{ref4} is the transmission probability, namely, 
\begin{equation}
T = {{(1+2\gamma)^2 (4-X^2)} \over {(1-2Q)^2 (4-X^2) + 4(P-QX)^2}}  ,
\label{eq18}
\end{equation}
where
\begin{equation}
P = 2z  ~,~ Q = z^2 - \gamma -\gamma^2 ,
\label{eq19}
\end{equation}
with
\begin{equation}
z= (\bar{\alpha}_j - \alpha)/ 2\beta ~,~ \gamma = (\bar{\beta}_j -\beta)/ 2\beta ~,~ j= p,m,o,
\label{eq20}
\end{equation}
being the reduced dimer-site energies and the intra-dimer coupling, respectively. The index $j$ is
chosen so as to represent the desired benzene type, using the appropriate pair of equations from
(\ref{eq11})-(\ref{eq16}).

\section{Results and Discussion}

The methodology of the two previous sections can now be implemented to produce the $T(E)$
curves for each of the three benzene types (para-, meta- and ortho-), when overlap is
ignored ($S=0$) and for several non-zero values of $S$. In the following calculations, the parameter
values chosen were the atomic-site energy $\alpha = 0$ and the bond energy $\beta =-0.5$, both
in the benzene molecules and their leads. As mentioned earlier, overlap $S$ is included only in the benzene
molecules, but not the leads, with its value varying from the no-overlap case of $S=0$ to a maximum of
0.25.

\begin{figure}[h]
\includegraphics[width=12cm]{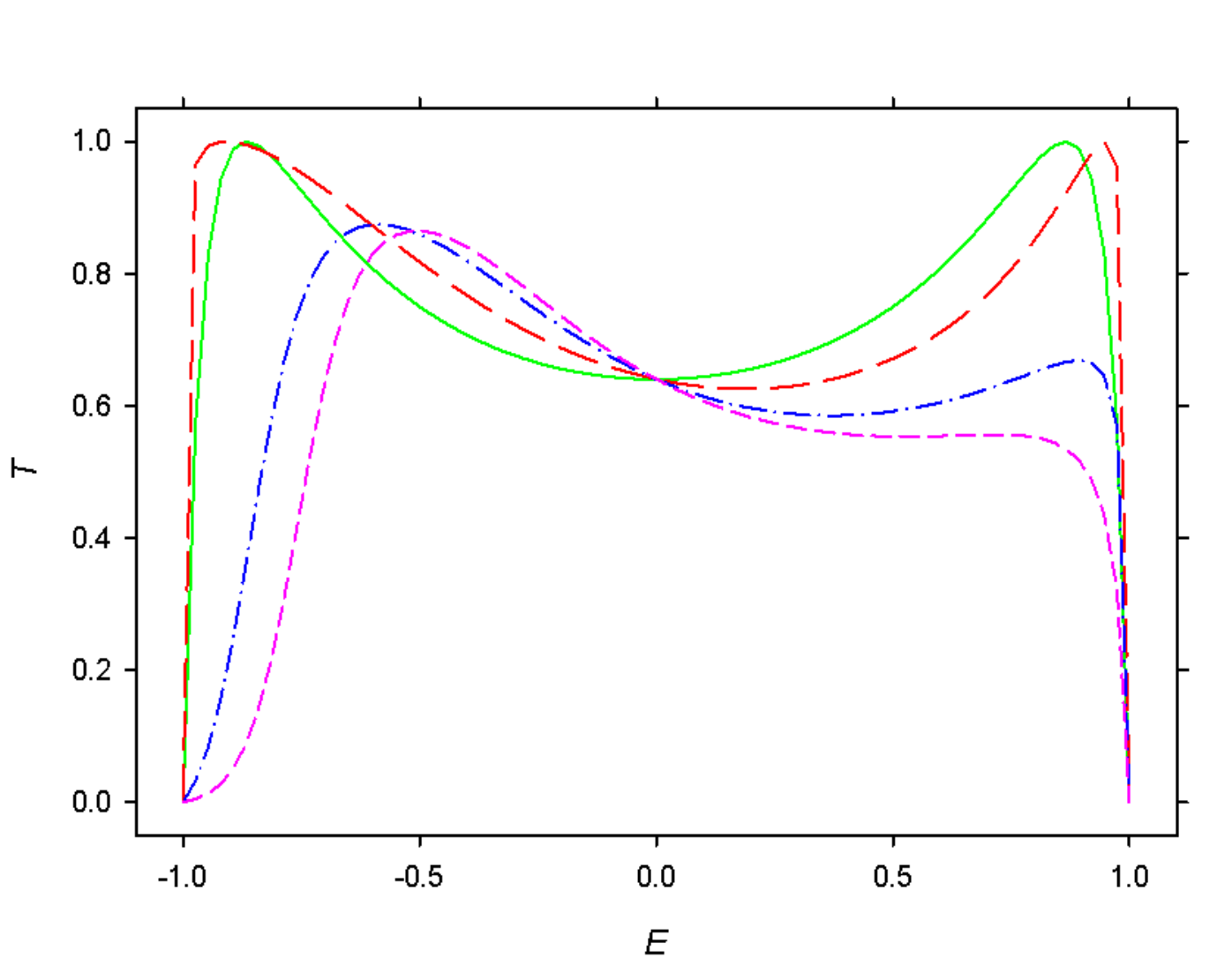}
\caption{Transmission $T$ versus energy $E$ for para-benzene, with overlap $S=$ (a) 0 (green solid curve), 
(b) 0.1 (red long-dashed), (c) 0.2 (blue dash-dotted), (d) 0.25 (pink short-dashed).}
\label{fig2}
\end{figure}

Figure \ref{fig2} shows the $T(E)$ curves for para-benzene, for the several values of $S$ indicated.
The no-overlap case, with $S=0$, is illustrated in Figure \ref{fig2}(a). The curve is symmetrical in
nature, with a pair of resonance maxima near the band edges, and a single local minimum at the band 
center $E=0$. Increasing $S$ from 0, even marginally, has the effect of destroying the symmetry of
 the curve, and diminishing the resonances, although small values of $S$ (see Figure \ref{fig2}(b)) 
 produce otherwise minor changes in $T(E)$. Larger increases in $S$ (see Figure \ref{fig2}(c) and (d)) 
 result in more significant changes, with further decreasing and broadening of the lower peak which 
 nonetheless maintains a clear maximum, but more
 noticeably for the upper peak, which undergoes some narrowing and, eventually, complete suppression.
 Thus, the asymmetry introduced by overlap shows a greater effect at higher energies, resulting in more
 favourable conditions for transmission at lower energies.
 
\begin{figure}[h]
\includegraphics[width=12cm]{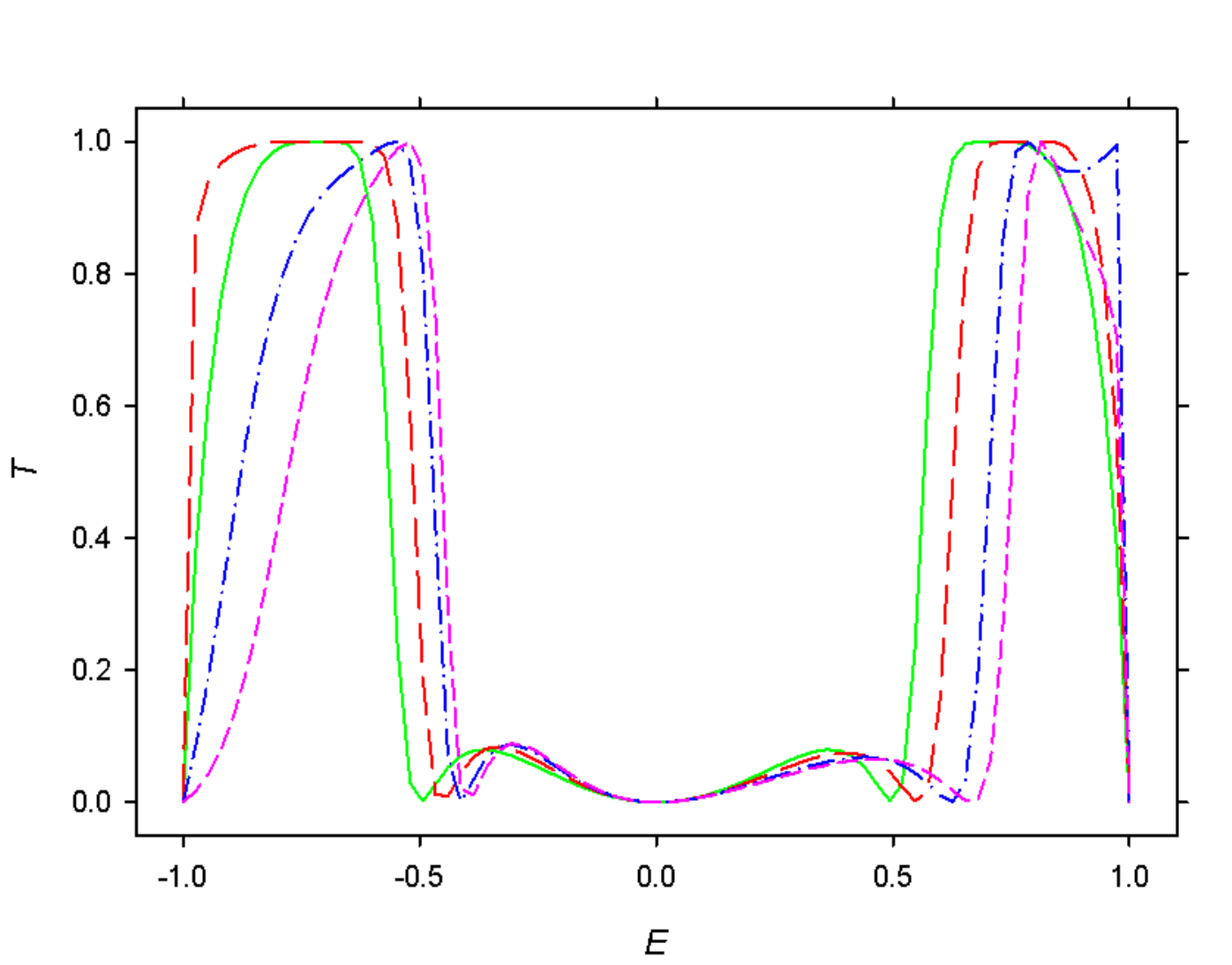}
\caption{Transmission $T$ versus energy $E$ for meta-benzene, with overlap $S=$ (a) 0 (green solid curve), 
(b) 0.1 (red long-dashed), (c) 0.2 (blue dash-dotted), (d) 0.25 (pink short-dashed).}
\label{fig3}
\end{figure}

 Turning next to meta-benzene, the case $S=0$ is shown in Figure \ref{fig3}(a). 
 The curve is symmetrical
 about the band center, with anti-resonances at $E=0$ and $E= \pm 0.5$. There are a pair of resonances
 at $E = \pm 0.71$, and a pair of smaller peaks at $E= \pm 0.37$, bounded by the anti-resonances. This
 structure, and in particular the anti-resonance at $E=0$, makes meta-benzene the poorest transmitter of the
 three benzene types \cite{ref5}. Taking $S$ to be non-zero (see Figure \ref{fig3}(b)-(d))  again removes
 the symmetry of the $T(E)$ curve, and all the more so with increasing values of $S$. Interestingly, and
 unlike the para-benzene case, the resonances are not lowered, but in fact, each is temporarily split into a pair of
 resonances (Figure \ref{fig3}(b)) before, with increasing $S$, the outer resonance of each pair is 
 gradually lowered until eventually disappearing (Figure \ref{fig3}(c) and (d)). Simultaneously, the lowermost
 peak broadens somewhat, while the uppermost narrows, due to the anti-resonances at $E= \pm 0.5$ shifting
 to slightly higher energies (0.66 and -0.4, in Figure \ref{fig3}(d)), while the band edges remain fixed, as does
 the anti-resonance at $E=0$. The two smaller peaks flanking $E=0$ are relatively unchanged, with the lower
 one increasing and the upper one decreasing marginally, and their widths changed somewhat by the shifting
 energies of the anti-resonances. As in the previous case, then, the greatest enhancement in $T(E)$ is shown
 at lower energies, while the least occurs at energies near $E=0$.
 
\begin{figure}[h]
\includegraphics[width=12cm]{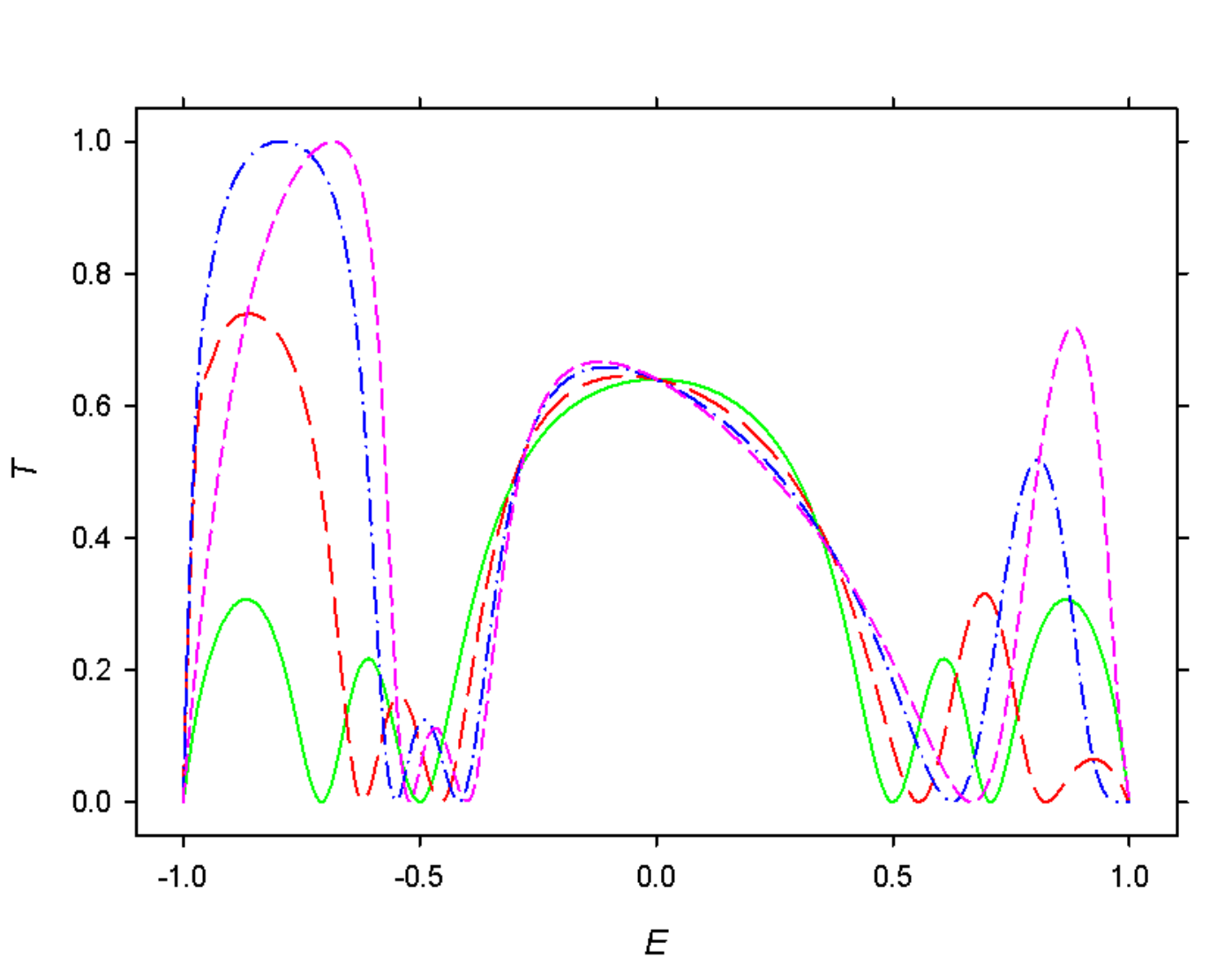}
\caption{Transmission $T$ versus energy $E$ for ortho-benzene, with overlap $S=$ (a) 0 (green solid curve),
(b) 0.1 (red long-dashed), (c) 0.2 (blue dash-dotted), (d) 0.25 (pink short-dashed).}
\label{fig4}
\end{figure}

 Finally, we consider ortho-benzene, starting from the case $S=0$, as shown in Figure \ref{fig4}(a). 
 Once again, the $T(E)$ curve is symmetrical, but with no resonances. There is a large central peak with a pair
 of smaller peaks on each side, separated by anti-resonances at $E=\pm 0.5$ and $E=\pm 0.71$. As in
 the two previous cases, increasing $S$ from 0 destroys the symmetry (Figure \ref{fig4}(b)-(d)), and indeed,
 more noticeably than in the other two cases. The change in the central peak is relatively mild, with the maximum
 increasing somewhat and moving to a slightly lower energy, as $S$ increases. But the real effect is in the side
 peaks, due partly to the anti-resonances shifting to higher energies, the upper duo more so than the lower pair.
 Indeed, the uppermost anti-resonance eventually coincides with the upper band edge, so that the uppermost
 sub-band disappears, all the while as its height is decreasing (Figure \ref{fig4}(c) and (d)). Its neighbouring peak
 increases substantially in height, while widening as well, to become one the graph's dominant features. 
 Meanwhile, the lowermost peak broadens and increases immensely to near-resonance status  (Figure \ref{fig4}(d)).
 On the other hand, its neighbouring peak is relatively unaffected, albeit lowered somewhat. Similar to the other
 two cases, $T(E)$ exhibits its biggest increases at lower energies.
 
 In comparing the effect of overlap in the three cases, it certainly appears that it is greatest for ortho-benzene,
 and smallest for meta-benzene. For even small non-zero values of $S$, such as 0.05 (not shown in the figures),
 the symmetry of the $T(E)$ curve is broken, but for such small $S$, the effect is otherwise slight, at least for
 para- and meta-. The effect increases steadily with increasing $S$. For para- and meta-, 
 the effect is greatest for $S \ge 0.15$, while it is always present for ortho-. In all cases, increasing $S$ has
 the general effect of shifting $T(E)$ to lower energies.

\section{Conclusion}

The rescaling technique has been extended, so as to incorporate overlap in the renormalization-decimation
equations. Subsequently, the method is applied to our previous work on transmission through benzene
molecules, so as to investigate the effect of overlap between the carbon atoms in benzene. The observed
effect is quite substantial, with the symmetry of the $T(E)$ curve being broken by even a small degree of overlap,
and increasing with $S$. Of the three types, the largest effect occurs for ortho-benzene and the smallest
for meta-benzene. These results are indicative that overlap should be included in future studies of benzene
transmission.

\section{Keywords}

Molecular electronics, benzene, overlap, renormalization-decimation technique, semi-empirical calculations


\begin{thebibliography}{99}

\bibitem{ref1} F. Chen, N.J. Tao, {\em Acc. Chem. Res.} {\bf 2009}, {\em 42}, 429.
\bibitem{ref2} R. McWeeny, {\em Coulson's Valence, 3rd ed.}, Oxford University Press, Oxford, {\bf 1979}.
\bibitem{ref3} T. Hansen, G.C. Solomon, D.O. Andrews, M.A. Ratner, {\em J. Chem. Phys.} {\bf 2009}, {\em 131}, 194704; G.C. Solomon, D.O. Andrews, T. Hansen, R.H. Goldsmith, M.R. Wasielewski, R.P. Van Duyne, M.A. Ratner, {\em J. Chem. Phys.} {\bf 2008}, {\em 129}, 054701.
\bibitem{ref4} R. Farchioni, G. Grosso, P. Vignolo, {\em Organic Electronic Materials}, (Eds. R. Farchioni, G. Grosso), Springer, Berlin, {\bf 2001}, pp. 89-125. 
\bibitem{ref5} K.W. Sulston, S.G. Davison, {\em arXiv} {\bf 2015}, 1505.03808.
\bibitem{ref6} B.A. Lippmann, J. Schwinger, {\em Phys. Rev.} {\bf 1950}, {\em 79}, 469; P.A. Mello,
N. Kumar, {\em Quantum Transport in Mesoscopic Systems}, Oxford University Press, Oxford, {\bf 2004}.



\end{thebibliography}
\end{document}